\documentstyle[12pt,amssymbols]{article}
\input math_macros.tex

\begin{document}
\begin{titlepage}
\begin{center}
\today     \hfill    LBL-37162\\

\vskip .5in

{\large \bf  Testing Single-Parameter Classical Standpoint Cosmology}
\footnote{This work was supported by the Director, Office of Energy
Research, Office of High Energy and Nuclear Physics, Division of High
Energy Physics of the U.S. Department of Energy under Contract
DE-AC03-76SF00098.}
\vskip .50in

{\bf{G.F. Chew}}

{\em Theoretical Physics Group\\
    Lawrence Berkeley Laboratory\\
      University of California\\
    Berkeley, California 94720}
\end{center}

\vskip .5in

\begin{abstract}
Experimental tests of homogeneous-universe classical
standpoint cosmology are proposed after presentation of conceptual
considerations that encourage this radical departure from the standard model.
Among predictions of the new model are standpoint age equal
to Hubble time, energy-density parameter
$\Omega_0 = 2 - \sqrt{2} =.586$, and  relations
between redshift, Hubble-scale distribution of matter and
galaxy luminosity and angular diameter.
These latter relations coincide with those of the standard model for zero
deceleration.
With eye to further tests, geodesics of the non-Riemannian standpoint metric
are
explicitly given.
Although a detailed thermodynamic ``youthful-standpoint'' approximation
remains to be developed (for particle mean free path small on standpoint
scale),
standpoint temperature  depending only on standpoint age is a natural
concept, paralleling energy density and redshift that
perpetuates thermal spectrum for cosmic background radiation.
Prospects for primordial nucleosynthesis are promising.

\end{abstract}
\end{titlepage}
\renewcommand{\thepage}{\roman{page}}
\setcounter{page}{2}
\mbox{ }

\vskip 1in

\begin{center}
{\bf Disclaimer}
\end{center}

\vskip .2in

\begin{scriptsize}
\begin{quotation}
This document was prepared as an account of work sponsored by the United
States Government. While this document is believed to contain correct
 information, neither the United States Government nor any agency
thereof, nor The Regents of the University of California, nor any of their
employees, makes any warranty, express or implied, or assumes any legal
liability or responsibility for the accuracy, completeness, or usefulness
of any information, apparatus, product, or process disclosed, or represents
that its use would not infringe privately owned rights.  Reference herein
to any specific commercial products process, or service by its trade name,
trademark, manufacturer, or otherwise, does not necessarily constitute or
imply its endorsement, recommendation, or favoring by the United States
Government or any agency thereof, or The Regents of the University of
California.  The views and opinions of authors expressed herein do not
necessarily state or reflect those of the United States Government or any
agency thereof, or The Regents of the University of California.
\end{quotation}
\end{scriptsize}

\vskip 2in

\begin{center}
\begin{small}
{\it Lawrence Berkeley Laboratory is an equal opportunity employer.}
\end{small}
\end{center}

\newpage
\renewcommand{\thepage}{\arabic{page}}
\setcounter{page}{1}
\noindent{\bf I. Introduction}

\medskip

Standpoint cosmology (Chew 1994, 1995), despite superficial phenomenological
similarity to the ``standard'' cosmology of Friedmann-Robertson-Walker (see
Weinberg 1972), differs profoundly in principle.
Standpoint cosmology is closer in spirit to ``kinematic cosmology'' (Milne
1935),
although a standpoint spacetime is {\it compact} with corresponding curvature.
Essential to both kinematic cosmology and to standpoint
cosmology is a concept of spacetime-localized ``big bang''
together with ``age'' measured
therefrom.
In the new model {\it age} belongs not to the entire universe as
 in the
standard model but rather to a ``standpoint'' where
``observer'' is
located.

Standard-model successes
(greater than those of kinematic cosmology where there is no curvature)
must eventually not only be matched but exceeded by the
new model if the latter is to survive.
The present paper, after reviewing conceptually-attractive novelties of
standpoint cosmology, displays explicitly in standpoint-based coordinate
systems
the homogeneous-universe geodesics.
Application thereof is then made (a) to the relation between standpoint age and
Hubble time, (b) to mean energy density, (c) to relation between redshift and
both luminosity distance and angular-size distance
and (d) to Hubble-scale distribution of matter.
Apart from the energy-density prediction $\Omega_0 = 2 - \sqrt{2}$, the
foregoing relations coincide not only with those of kinematic cosmology but
with
those  of the standard model for zero ``deceleration''.

A detailed thermodynamic approximation remains to be developed.
It will nevertheless become plausible from what follows that, when particle
mean
free path is small on the (Hubble) scale of some standpoint, a standpoint
temperature can be defined that depends only on standpoint age
and that decreases as age advances.
Age-temperature correlation dovetails with a photon redshift controlled
entirely
by ratios of standpoint ages.
We shall be led to qualitative  understanding of cosmic background-radiation
and to optimism about nucleosynthesis within standpoint cosmology.
The new model leaves undisturbed the theory of fluctuations, on length scales
small compared to Hubble scale, that arise from weak Einstein gravity (Chew
1995).
\newpage

\noindent{\bf II. Conceptual Novelties}

\medskip

The new model is economical; a {\it single} standpoint-associated parameter
of length dimensionality, designated $R$, controls ``radius
of universe'' (seen from standpoint) together with standpoint age ($c=1$) and
Hubble time.
  As is the case for Milne's cosmology,
there is no scale parameter depending on universal time, no deceleration
parameter, no cosmological constant.
In tandem with the gravitational constant $G$, the parameter $R$ determines
mean
energy density.
Paucity of parameters places the new model in immediate jeopardy of
experimental
falsification.

As in Milne's cosmology, there is no meaning for universe beyond a horizon tied
to big bang.
All matter is causally connected- -sharing a spacetime-localized big-bang
origin.
Only optical opacity obstructs observation from any standpoint of the entire
classical universe.
Nevertheless there is a sense in which the universe is ``infinite'': departing
from
some standpoint in a fixed spatial direction, there is no limit to the
different
standpoints of same age to be encountered.
In any standpoint coordinate system a huge quantity of matter concentrates near
horizon.
In Mach spirit one may think of such ``maximally-distant'' matter as
responsible
for the Minkowski metric tensor that holds sway (in homogeneous-universe
approximation) near any standpoint in that standpoint's coordinate system (see
Chew 1995).

Despite the prevailing physics paradigm of covariance within a {\it unique}
unbounded spacetime, the new model attributes to {\it each} standpoint a {\it
separate} compact spacetime endowed with a special set of coordinates.
This coordinate system is suitable for describing experiments carried out in
the
neighborhood of that standpoint.
On the scale of $R$ (Hubble scale), ``homogeneous universe'' presents the same
appearance from any standpoint when described by the coordinates belonging to
that standpoint.
Only a {\it portion} of one compact standpoint spacetime generally maps onto
another such spacetime.
It will nevertheless be shown that familiar Poincar\'e symmetry (of a {\it
unique} spacetime) prevails (approximately) within neighborhoods that are small
on Hubble scale.

The separate compact spacetimes are tied together by invariant metric combined
with common origin of coordinate systems.
The common origin is identified with ``big bang''.
A ``newly-born'' standpoint originates in big bang and moves ``outward'' in
{\it any}
``old'' standpoint coordinate system along a well-defined positive-timelike
geodesic.
Standpoint age is proportional to invariant ``distance'' from big bang.
Each standpoint trajectory being labelable by ``initial velocity'' (near big
bang), any standpoint is specified by age plus initial velocity.
Because all standpoint spacetimes are Minkowskian near big-bang origin,
``homogeneity of universe'' corresponds unambiguously to a nonintegrable
Lorentz-invariant distribution of initial standpoint velocities.
Nonintegrability amounts to the previously-emphasized ``infinite universe''.

Standpoint-spacetime metrics are generally non-Riemannian, although they
approach Minkowski form not only near big bang but, in homogeneous-universe
approximation, also near standpoint.
The metric is Riemannian for {\it radial} homogeneous-universe motion in any
standpoint coordinate system and for general motion near standpoint if
inhomogeneity is ``weak'' (see Chew 1995).
In the latter case, Einstein theory of gravity applies in standpoint
neighborhood (small on Hubble scale).
Near ``strong'' inhomogeneities (``black holes''), the
not-yet-understood non-Riemannian character of standpoint metric becomes
important.

``Standpoint'' represents separation between past and future - - i.e., the
``present''.
Metric describing the past is different from that describing the future when
Hubble-scale times are considered.
Only for time displacements from the present short on Hubble scale is there
(approximate) equivalence.

Many conventionally-tolerated displacements are disallowed in a compact
standpoint spacetime.
Consistency depends on additivity of {\it positive} timelike or lightlike
displacements associable with matter motion.
Asymmetry between past and future is dramatically manifested by an impassible
future boundary - - called ``abyss'' (Chew 1994).
Prediction of future based on present measurements - - i.e., measurements made
near standpoint - - cannot extend beyond this boundary.
The abyss limitation correlates with geodesics and may be regarded in the
spirit
of ``Schwartzschild radius''
accompanying a mass of order $\rho R^3$, where $\rho$ is energy density at
standpoint.

The only region within a standpoint spacetime accessible to {\it measurement}
is
the neighborhood of the standpoint's backward light cone.
The remainder of a standpoint spacetime facilitates {\it prediction} of results
from (future) measurements to be carried out near older standpoints and {\it
verification}
of prediction based on (earlier) measurements made near younger standpoints.
Essential to the integrity of standpoint cosmology's emphasis on measurement
correlation is the ``stability'' of lightlike geodesics: a lightlike geodesic
in
one standpoint spacetime maps onto lightlike geodesic within any other (where
mapping is possible).
Classical-measurement correlation dovetails with $S${\it -matrix}
interpretation of {\it quantum} standpoint cosmology (Chew 1994).

Although the present paper will not discuss the quantum underpinning of
standpoint cosmology, here {\it defining} the classical model by the metric of
standpoint spacetime, this metric was uniquely
inferred from symmetry properties of a
more fundamental quantum model of expanding universe.
Only for standpoints whose $R$ greatly exceeds $R_{min} \sim$ 10 cm (age large
in nanoseconds) does 3-space in the quantum model achieve classical
significance.
Quantum-model meaning for ``location'' within a standpoint spacetime arises in
conjunction with meaning for ``particles''. In a ``dense'' region of the
universe - - where $R \ll R_{min}$ - - neither particles nor 3-space enjoy
model
meaning.
According to the quantum model, ``diluteness'' is essential to classical
significance for 3-space.

A semantic observation: although classical standpoint cosmology, with
underpinning that lacks {\it a priori} spacetime, fails to accord with all
aspects of general relativity, the model considered here may be described as
``more relativistic'' than the standard model.
The latter, after all, is characterized by a universal time.

Milne's 1935 cosmology corresponds to standpoints of {\it infinite }
age, which  have past but
no present and  no future.
It often turns out calculationally convenient to invoke infinite age
where the metric is Minkowskian, but {\it physical}
spacetime belongs to a {\it present} where the surrounding  spacetime is
curved.

\newpage

\medskip

\noindent{\bf III. Specially-Coordinated Standpoint Spacetimes}

\medskip

Because the spacetime belonging to a standpoint is compact, with boundary and
well-defined ``center'', there is an accompanying natural system of
coordinates.
A standpoint locates at the center of its {\it own} spacetime where it is ``at
rest''.
In coordinate systems other than its own, a standpoint is displaced from center
and generally is in motion.
Any (compact) standpoint spacetime may be described as the intersection of
interiors of forward and backward light cones whose vertices share the
standpoint's spatial location while each vertex locates an interval $R$ in time
from the standpoint, one vertex in the standpoint's past and the other in its
future. (The past vertex is identifiable with big bang.)

Using the boldface symbol ${\bf{R}}$ to designate a standpoint and the 4-symbol
${\bf{x}}_{\bf{R}} = ( t_{\bf{R}}, \vec{r}_{\bf{R}})$ for the special attached
coordinates, restriction to the double-cone interior amounts to coordinates
being constrained to the interval,
$$
0 \ \leq \ t_{\bf{R}} \pm \abs{\vec{r}_{\bf{R}}} \ \leq \ 2R.\eqno(III.1)
$$
The ${\bf{R}}$ standpoint locates at $t_{\bf{R}} = R, \vec{r}_{\bf{R}} =0$,
i.e.,
 at the double-cone center.
(Big bang locates at $t_{\bf{R}}=0, \vec{r}_{\bf{R}}=0$).
It will be seen in Section IV that standpoint-spacetime geodesics
curve in  conformity to (III.1) - - matter inside the double cone being
  unable to cross the boundary.
This curvature constitutes a major departure from Milne's 1935 kinematic
cosmology.

Portions of one standpoint spacetime map onto portions of others.
Explicit mapping rules (in homogeneous-universe approximation) will be
presented.
Mappings are anchored by big bang - - the origin of one coordinate system
mapping onto the origin of any other and, because all spacetimes are
asymptotically Minkowskian in neighborhood of origin $(t_{\bf{R}} \ll R)$,
 the (infinitesimal) positive timelike or lightlike 4-vectors
${\bf{x}}_{\bf{R}}$
 are there related to each other by Lorentz boosts.
A convenient corollary is explicit elaboration of the symbol ${\bf{R}}$ into
the
4-symbol $(R, \vec{\beta})$, with the 3-vector $\vec{\beta}$ interpretable as
``initial rapidity'' of standpoint.
That is, in the coordinate system belonging to a zero-rapidity standpoint
${\bf{R}} = (R, \vec{0})$, some (other) ``very young''
standpoint located at ${\bf{x}}_{\bf{R}}$ (with $t_{\bf{R}}\ll R)$ has rapidity
$\vec{\beta}$ such that
$$
\tanh
\abs{ \vec{\beta} } = { \abs{ \vec{r}_{\bf{R}} }\over t_{\bf{R}} }, \ \
{\vec{\beta}\over\abs{\vec{\beta}} } \ = \
{ \vec{r}_{\bf{R}}\over \abs{ \vec{r}_{\bf{R}}
 }}.\eqno(III.2)
$$
We shall see that, as this standpoint of initial -rapidity $\vec{\beta}$
grows to an age of order $R$, its rapidity in the $(R, \vec{0})$ coordinate
system diminishes so as to keep the moving standpoint within the compact $(R,
\vec{0})$ spacetime.
This deceleration, gravitationally interpreted, will in Section IV determine
mean energy density in terms of $R$ and $G$.

Mapping between $(R, \vec{\beta})$ and ($R', \vec{\beta}')$ coordinates is
conveniently achievable by a 3-step process involving standpoints of {\it
infinite} age:
$$
(R, \vec{\beta}) {\stackrel{A}{\longrightarrow}} (\infty, \vec{\beta})
{\stackrel{B}{\longrightarrow}}\  (\infty, \vec{\beta}')
{\stackrel{C}{\longrightarrow}}\
(R', \vec{\beta}')\eqno(III.3)
$$
Step B we shall see to be a simple Lorentz boost (with counterpart
in kinematic cosmology).
    Steps A and C at fixed initial rapidity  are also simple transformations
but of a
completely different type exposed in Section IV after standpoint-spacetime
metric is introduced.
Fixed-$\vec{\beta}$ mappings between coordinate systems of different ages are
generally defined only for {\it portions} of the involved spacetimes.

\newpage
\noindent {\bf IV. Geodesics}

\medskip

Compactness of standpoint spacetime, accompanied by non-Riemannian metric (Chew
1994), precludes applicability of numerous notions from general relativity.
Surviving, nevertheless, is representation of gravity through matter motion
along geodesics; gravitational mass continues to be indistinguishable from
inertial mass.
Classical metric is controlled by the symmetry of an underlying quantum
dynamics whose description here is impractical.
A convenient consideration is that {\it radial}
homogeneous-universe motion in a standpoint spacetime is describable by a
quadratic (Riemannian-like) form.
For radial displacements with respect to ${\bf{R}}$ standpoint, an increment of
``distance'' turns out to be  given by
$$
ds^2 = \{(1- t_{\bf{R}}/2R)^2 - (r_{\bf{R}}/2R)^2\}^{- 1/2} (dt^2_{\bf{R}} -
dr^2_{\bf{R}}),\ r_{\bf{R}}\equiv \abs{\vec{r}_{\bf{R}}},\eqno(IV.1)
$$
even though nonradial motion requires a {\it quartic} form.
(Absence of subscript on $ds^2$ is remindful of distance invariance under
change
of standpoint.)
The radial metric (IV.1) will  generate the required mappings between
standpoint spacetimes of same rapidity but different $R$.
We shall not here need the quartic expression of more general metric.

Notice that the radial metric (IV.1) is singular along the backward-light-cone
(future) spacetime boundary where $r^2_{\bf{R}} = (2R-t_{\bf{R}})^2$.
This singularity, present also in the general metric, prevents
any geodesic from penetrating the future boundary - - which has been called
``abyss'' (Chew 1994). Notice further that in big-bang neighborhood (i.e.,
$t_{\bf{R}}\ll R)$ or, equivalently, in the limit $R \to \infty$, the
anticipated Minkowskian form is achieved.
In standpoint neighborhood
$(\abs{t_{\bf{R}}-R} \ll R, r_{\bf{R}} \ll R)$
the metric also is Minkowskian although here
$ds^2 = 2 (dt^2_{\bf{R}} - dr^2_{\bf{R}})$.
The factor 2 will be found below to influence standpoint age.

The metric (IV.1) implies the radial equation of motion (geodesic differential
equation)
$${ d^2 r_{\bf{R}}\over dt^2_{\bf{R}} }=- \half \left[ 1- \left({dr_{\bf{R}}
\over
 dt_{\bf{R}} }\right)^2\right] {r_{\bf{R}} + (2R-t_{\bf{R}}) {dr_{\bf{R}}\over
dt_{\bf{R}}}
 \over (2R-t_{\bf{R}})^2 - r^2_{\bf{R}} },\eqno(IV.2)
$$
for which explicit solutions will below be presented.
Because radial motion with respect to one  standpoint maps onto nonradial
motion
with respect to another standpoint of different spatial location (different
initial
rapidity), the mapping strategy (III.3) generates
from solutions to (IV.2) the most general
homogeneous-universe geodesics.

A Newtonian-gravitational interpretation of the linear approximation to (IV.2)
in standpoint neighborhood, i.e., of the approximate equation of motion
$$
{d^2r_{\bf{R}}\over dt^2_{\bf{R}} } \approx - \half \left( {r_{\bf{R}}\over
R^2}
+ {1\over R} {dr_{\bf{R}}\over dt_{\bf{R}} }\right),\eqno(IV.3)
$$
allows an inference of mean energy density in standpoint 3-space.
 At time $t_{\bf{R}} =R$,
consider matter spatially displaced from ${\bf{R}}$-spacetime center (i.e.,
from
${\bf{R}}$ standpoint) by a distance $r_{\bf{R}}$ that is small compared to
$R$.
Let this matter be at rest with respect to that standpoint - - displaced
slightly
from ${\bf{R}}$ - - which coincides in location with the matter.
It may be deduced from formulas in Section VI that such
``stationary''  matter has radial velocity
in the ${\bf{R}}$
system,
$$
{dr_{\bf{R}}\over dt_{\bf{R}} } = {1\over \sqrt{2}} \ {r_{\bf{R}}\over {\bf{R}}
} +
\ \hbox{order} \ \left( {r_{\bf{R}}\over {\bf{R}} }\right)^2.\eqno(IV.4)
$$
It then follows from (IV.3) that nonrelativistic matter acceleration in the
${\bf{R}}$
system, in the neighborhood of ${\bf{R}}$ standpoint, is
$$
{d^2 r_{\bf{R}}\over dt^2_{\bf{R}} }=- \half \left(1 + {1\over \sqrt{2} }
\right) \
{r_{\bf{R}}\over R^2}
+ \ \hbox{order} \ \left({r^2_{\bf{R}}\over R^3}\right).\eqno(IV.5)
$$

In Newtonian terms the foregoing acceleration is attributable to a restoring
gravitational
force that resists displacement from the center of a spherically-symmetric mass
distribution (whose radius is of order $R$.)
If mass density at center is $\rho_{\bf{R}}$, the Newtonian gravitational
 potential at small
$r_{\bf{R}}$ is
$$
G \ { {4\pi\over 3} r^3_{\bf{R}} \rho_{\bf{R}}
\over r_{\bf{R}} },\eqno(IV.6)
$$
corresponding to an acceleration (toward the center)
$$
- G \ {8\pi\over 3} r_{\bf{R}} \rho_{\bf{R}}.\eqno(IV.7)
$$
Equating (IV.7) with (IV.5) yields
$$
\rho_{\bf{R}} = {3\over 16\pi} \left(1+ {1\over \sqrt{2}} \right)\  {1\over
GR^2}.\eqno(IV.8)
$$
Once $R$ has below been related to Hubble time, it will be found that (IV.8)
corresponds to the conventionally-defined density parameter (fraction of
``critical''
density in standard model),
$$
\Omega_0 = 2 - \sqrt{2} = .586.\eqno(IV.9)
$$

The foregoing prediction of $\Omega_0$ is provisional, subject to systematic
derivation
of classical standpoint cosmology as a dilute-universe approximation to the
more exact quantum
model.
Such a derivation would relate $G$ to a ``more-fundamental''
small dimensionless parameter (Chew, 1994).
In the interim, before a quantum-based theory of gravity becomes available,
we are leaning on experimentally-supported (sub-Hubble-scale) features
of classical Newton-Einstein theory (see Chew 1995) where $G$ is regarded as a
fundamental
constant of nature.

The structure of (IV.1) exemplifies the general principle that the limit $R \to
\infty$ for fixed ${\bf{x}}_{\bf{R}}$ leads to Minkowskian metric.
In this (Milne)  limit the spacetime becomes {\it noncompact} and {\it unique}
for all
$\vec{\beta}$ - - corresponding to the forward light cone with big bang as
vertex.
Infinite-$R$ coordinate systems, each labeled by a 3-vector rapidity, all
describe the {\it same} spacetime.
These systems are related to each other by Lorentz transformations, with
${\bf{x}}_{\infty, \vec{\beta}}=(t_{\infty, \vec{\beta}}, \vec{r}_{\infty,
\vec{\beta}})$
behaving as a 4-vector. (Poincar\'e displacements are {\it{not}} allowed.)
Infinite-$R$ spacetime, while extremely useful as intermediary in the mapping
strategy (III.3), is not a {\it physical} spacetime.
``Usual physics'' situates in the neighborhood of some finite-$R$ standpoint
and
is to be described by the attached coordinate system.
Section IX will explain how usual Poincar\'e symmetry (under displacements as
well as Lorentz transformations) prevails {\it approximately} within standpoint
neighborhoods small on the scale of $R$.

Invariance of the radial distance given by (IV.1) implies the fixed-rapidity
mapping, $(R, \vec{\beta})\to(\infty, \vec{\beta})$,
$$
t_{\infty} \pm r_{\infty} = 4R\left\{1-\sqrt{1-{t_{\bf{R}}\pm r_{\bf{R}}\over
2R}}\right\}, \ \ {\vec{r}_\infty\over r_\infty} = {\vec{r}_{\bf{R}}\over
r_{\bf{R}}},\eqno(IV.10)
$$
with the inverse, $(\infty, \vec{\beta}) \to (R, \vec{\beta}),$
$$
t_{\bf{R}} = t_\infty - {t^2_\infty + r^2_{\infty}\over 8R},\eqno(IV.11a)
$$
$$
\vec{r}_{\bf{R}} = \vec{r}_{\infty} (1- {t_\infty\over 4R}).\eqno(IV.11b)
$$
Here the rapidity index $\vec{\beta}$ has been suppressed.
The interval $0\leq t_{\bf{R}} \pm r_{\bf{R}} \leq 2R$
is mapped onto the interval
$0\leq t_\infty \pm r_\infty \leq 4R$ and vice versa.
Straightline geodesics in infinite-$R$ coordinates transform into curved
geodesics in finite-$R$ coordinates.
(When these latter geodesics are radial, they satisfy the differential equation
(IV.
2).)
The most general geodesic may be written in infinite-age coordinates as the
straight line
$$
\vec{r}_{\infty} = \vec{a} + \vec{b} (t_\infty -c), \eqno(IV.12)
$$
with $\vec{a}, \vec{b}, c$ a set of 7 constants constrained by
$c \ \geq 0,\ \abs{ \vec{a}} \leq c,\ 0 \leq \abs{\vec{b}} \leq 1$.
Here the 4-vector ${\bf{x}}_0 = (c, \vec{a})$ locates ``source'' of matter
trajectory while the 3-vector $\vec{b}$ is matter velocity.
The special geodesics followed by standpoints correspond to
${\bf{x}}_0 =0$ with
$\abs{\vec{b}}=\tanh\abs{\vec{\beta}}$ and $\vec{b}/\abs{\vec{b}}
= \vec{\beta}/\abs{\vec{\beta}}$.
\newpage

\noindent{\bf V. Standpoint Age}

\medskip

What time registers on a clock carried by an observer who starts clock close to
big bang and moves along a standpoint trajectory?
The clock adds up time increments $dt_{\bf{R}}$ in a succession of {\it
different}
coordinate systems as $R$ increases, with the relation
$$
dt_{\bf{R}} = {1\over \sqrt{2}} ds \eqno(V.1)
$$
prevailing continuously along the trajectory. It follows that ``standpoint
age''
is ${1\over \sqrt{2}}$ times its invariant distance from big bang.
Recognizing $\vec{r}_\infty$ to be zero everywhere along trajectory, distance
from big bang is
$$
s = \left( t^2_\infty - r^2_\infty\right)^{1/2} = t_\infty. \eqno(V.2)
$$
{}From (IV.10) one calculates
$$
t_\infty (t_{\bf{R}} = R, \ r_{\bf{R}} =0) = 4R\ \left(1-
{1\over\sqrt{2}}\right),\eqno(V.3)
$$
so standpoint age is
$$
\eqalignno{
\tau_{\bf{R}} &= {1\over \sqrt{2} }\ 4R\ \ \left(1-{1\over \sqrt{2}}\right),\cr
       &= {R\over {1\over\sqrt{2}} + \half}.&(V.4)\cr}
$$
Phenomenologically, what we are calling ``standpoint age'' is the quantity
commonly called ``age of universe''.
The latter terminology, which fits the standard model, seems inappropriate here
and we shall avoid it.

\newpage

\noindent{\bf{ VI. Redshift and Hubble Parameter}}

\medskip

The outcome of the following calculation of redshift is so simple that we state
it immediately.
The redshift factor commonly denoted $1+z$ is equal to the ratio of observer
age
to source age (or observer$-R$ to source$-R$). Equivalently,
$
1+z =e^\Delta,
$
where $\Delta$ is the magnitude of source-standpoint {\it initial} rapidity
when
observer-standpoint (initial) rapidity is zero.
The simplicity of this relation raises expectation of a transparent derivation.
Unhappily we are not presently in possession of such.
The calculation to follow combines Doppler redshift due to source motion in
observer system with ``propagation redshift'' due to gravity experienced by
photons moving through observer-standpoint spacetime.

{}From (IV.11) and (IV.12) with ${\bf{x}}_0=0$, it is straightforward to
calculate in
observer system the radial rapidity $\Delta_s$ of a source, located on the
observer's backward light cone, that follows the trajectory of a standpoint
whose {\it initial} rapidity magnitude was $\Delta$.
One finds
$$
\Delta_s = \half \ln \left({e^{3\Delta}\over e^{-\Delta} +
2^{3/2}\sinh\Delta}\right).\eqno(VI.1)
$$
It may be verified that there is deceleration - - i.e. $\Delta_s < \Delta$.
(For small $\Delta, \Delta_s \approx (2 - \sqrt{2})\Delta$.)
The Doppler redshift factor is then
$$
e^{\Delta_s} = {e^{3/2\Delta}\over (e^{-\Delta}+2^{3/2}\sinh\Delta)^{1/2}}.
\eqno(VI.2)
$$

What about propagational redshift?
Here we need to study geodesics along the observer's backward lightcone.
{}From a computation described in the Appendix  one finds a propagational
redshift
factor
$$
\left( 1 + {2r_s\over R}\right)^{1/4},\eqno(VI.3)
$$
where $r_s$ is distance to source in observer system, the parameter $R$
belonging to observer standpoint.
The distance to source, from a calculation paralleling that leading to (VI.1),
turns out to be
$$
r_s =
e^{-2\Delta}\sinh\Delta(\cosh\Delta+\sqrt{2}\sinh\Delta)\tau_R,\eqno(VI.4)
$$
and, remembering (V.4), one then calculates from (VI.4) that
$$
\left(1+{2r_s\over R}\right)^{1/2} =
e^{-\Delta}(e^{-\Delta}+2^{3/2}\sinh\Delta).\eqno(VI.
5)
$$
Thus the product of (VI.3) with (VI.2) is simply $e^\Delta$.

That $e^\Delta$ gives the ratio of observer age to source age follows from the
mapping of observer backward light cone onto a corresponding backward light
cone
in the infinite-age coordinate system that has spatial origin at observer.
Along this infinite-age cone,
$$
t_\infty + r_\infty = s^{observer},\eqno(VI.6)
$$
and remembering the general relation $s=(t^2_\infty - r^2_\infty)^{1/2}$,
together with the special relation tanh
 $\Delta = r^{source}_\infty/t^{source}_\infty$, so
$$
\eqalignno{
t^{source}_\infty &= s^{source}\cosh\Delta,\cr
r^{source}_\infty &= s^{source}\sinh\Delta,&(VI.7)\cr}
$$
it follows from (VI.6) that
$$
e^\Delta s^{source} = s^{observer}.\eqno(VI.8)
$$

Consider next the Hubble parameter, {\it{phenomenologically}} definable as
$$
H_0 \equiv \lim_{r_s\to 0} {z\over r_s}.\eqno(VI.9)
$$
Because Formula (VI.4) exhibits a linear relation between $r_s$ and $\Delta$
for small $\Delta$,
$$
\lim_{\Delta \to 0} {r_s\over \Delta} = \tau_R,\eqno(VI.10)
$$
while in the same limit $z/\Delta \to 1$, it follows that
$$
H_0 = \tau^{-1}_R.\eqno(VI.11)
$$

Before closing this section we remark that, according to (VI.4), the upper
limit
of $r_s$ - - distance to source located on standpoint backward light cone - -
is
$R/2$, reached as $\Delta \to \infty$.
In other words, $R/2$ is ``radius of the ${\bf{R}}$ standpoint's universe.''
Such a statement, as emphasized above in Section II, can be misleading inasmuch
as Section VIII will show that (apart from quantum limitation) an
indefinitely-large amount of matter concentrates near standpoint horizon.
Classically speaking, our universe  is infinite.
\newpage

\noindent{\bf{VII. Luminosity Distance}}

\medskip

In this section we shall compute luminosity distance (as defined by Weinberg
1972) and will find
$$
d_L(z) = \tau_{\bf{R}}(z + z^2/2).\eqno(VII.1)
$$
Although this result coincides with the standard-model formula for zero
deceleration parameter, an independent derivation is required.
There is no present understanding of the coincidence.

We are concerned with observer-system trajectories followed by photons
emitted isotropically from the spatial origin of the source coordinate system.
Let us designate by $\theta_s$ the angle of emission in source system with
respect to the direction,
$
\vec{n} \equiv - \vec{\Delta}/\Delta,$
that (in either system) connects source to observer.
Our task is to compute, for extremely small $\theta_s$, the photon impact
parameter with respect to observer in observer system; this impact parameter
will be equated with $\theta_s$ times ``effective distance''.
After attention to redshift loss of photon energy and to extension of observer
time interval during which some collection of photons is received, ``luminosity
distance'' will emerge.

The direction $\vec{n}'$ of photon emission in the source coordinate system
$(\cos \theta_s = \vec{n}'\cdot\vec{n})$ is also the direction of photon
propagation in infinite-age rapidity-$\vec{\Delta}$ coordinates.
In the latter system photon spacetime location is given by the 4-vector
${\bf{x}}_{\infty,\vec{\Delta}}$ which we abbreviate by ${\bf{x}}'= (t',
\vec{{r}}\ ')
$.
Introducing photon distance from big bang
$$
s = (t'^2 - r'^2)^{1/2},\eqno(VII.2)
$$
it is convenient to represent photon trajectory as a 4-vector function of
 the invariant $s$, which at emission takes the value $s^{source}$ and at
observation equals $s^{observer}$.
Under-$\vec{\Delta}$ boost the 4-vector ${\bf{x}}'$ transforms to the 4-vector
${\bf{x}} = {\bf{x}}_{\infty, \vec{0}} = (t, \vec{r})$
that locates photon in {\it zero-rapidity} infinite-age coordinates.
Invariance of $s$ means $s=(t^2-r^2)^{1/2}$.
When the photon is near observer, $t\approx s^{observer}$ and
$\vec{r} \approx 0$; near observer it follows that
$$
r' \approx s^{observer} \sinh \Delta.\eqno(VII.3)
$$

Employing the symbol ${\bf{R}}$
for observer standpoint, with the coordinates $(t_{\bf{R}}, \vec{r}_{\bf{R}})$
{\it physically} locating the photon, we seek for photon near observer
the component of $\vec{r}_{\bf{R}}$ {\it transverse} to $\vec{\Delta}$.
Because at observer
$$
t=s^{observer}=4R\left(1-{1\over\sqrt{2}}\right),\eqno(VII.4)
$$
Formula (IV.11) tells us that, near observer,
$$
\vec{r}_{\bf{R}} \approx {1\over \sqrt{2}}\vec{r}.\eqno(VII.5)
$$
Using the subscript ``$tr$'' to denote transverse components of 3-vectors,
it follows that the desired impact parameter is
$$
r_{{\bf{R}},tr}  \approx  {1\over\sqrt{2}} \ r_{tr},\eqno(VII.6)
$$
for $s=s^{observer}$.
Because a Lorentz boost does not alter the transverse component of a 4-vector,
we have
$$
\vec{r}_{tr} = \vec{r}\ '_{tr},\eqno(VII.7)
$$
and consequently
$$
r_{{\bf{R}},tr}\approx {1\over\sqrt{2}}\  r'_{tr}.\eqno(VII.8)
$$
Finally, because all trajectories are straight lines in infinite-age
coordinates,
it follows that
$r'_{tr} \approx \theta_s r'$, and Formula (VII.3) together with (VII.8) leads
to
$$
r_{{\bf{R}},tr}\approx \theta_s \tau_{\bf{R}} \sinh\Delta.\eqno(VII.9)
$$
For $\Delta \ll 1$ where, according to (VI.4), $\tau_{\bf{R}}\Delta$
approximates
(observer-measured {\it or} source-measured)
distance between source and observer, the result (VII.9) agrees with
straightline photon propagation through a unique flat space; but for $\Delta
\gtap 1$, (VII.9) becomes drastically non-Euclidean.
(As $\Delta \to \infty, \tau_{\bf{R}} \sinh \Delta \to \infty$ whereas distance
to
source approaches $R/2$.)

Luminosity is source-generated energy received at observer per unit area per
unit
time.
Impact parameter deals with photons per unit transverse area although not with
energy per unit time.
Momentarily deferring the latter, we recognize $\theta^2_s/4$ to be the
{\it{fraction}}
of photons isotropically emitted in source system that eventually arrive
within the impact parameter (VII.9).
Because the observer-system transverse area in question is $\pi \theta^2_s
d^2_e
(\Delta)$, where
$$
d_e(\Delta) \equiv \tau_{\bf{R}} \sinh\Delta,\eqno(VII.10)
$$
the fraction of photons eventually arriving per unit area at observer is $[4\pi
d^2_e (\Delta)]^{-1}$.
Geometrically speaking, therefore, $d_e(\Delta)$ acts as ``effective
distance''.

However, fraction of {\it energy} emitted per unit source time that is received
{\it per unit observer time} is reduced by a factor $e^{-2\Delta}$ - - redshift
reduction of photon energy being by a factor
$e^{-\Delta}$,
with a second
factor $e^{-\Delta}$ arising from the ratio between source-time interval for
emission of a number of photons and receiver-time interval for reception of
these photons. Following Weinberg 1972, if instead of (VII.10) we define
``luminosity distance'' by
$$
d_L(\Delta) \equiv \tau_{\bf{R}} e^\Delta \sinh\Delta,\eqno(VII.11)
$$
then $[4\pi d^2_L(\Delta)]^{-1}$ gives ``luminosity fraction'' per unit area at
observer.
Remembering that $e^\Delta = 1+z$, we may rewrite (VII.11) as
$$
d_L = \tau_{\bf{R}} (z + \half z^2),\eqno(VII.12)
$$
finally achieving the result advertised above in (VII.1).

With {\it inversion} of source and observer, the calculation leading to (VII.9)
yields the angle subtended at observer (in observer coordinates) by a source
diameter (in source coordinates). The result is equivalent to
``angular-diameter
distance'' (Kolb, Turner  1990)

$$
\eqalignno{
d_A(\Delta)=\tau_{source}\sinh \Delta &= e^{-2\Delta} d_L (\Delta)\cr
&= \tau_{\bf{R}} {z(1+ z/2)\over (1 + z)^2}.&(VII.13)\cr}
$$
Formula (VII.13) agrees with that given by the standard model with
zero deceleration.
Note that, according to (VII.13), the observed subtended angle approaches a
{\it constant} $(2
{d_{source}\over \tau_{\bf{R}}}) $ as $\Delta \to \infty$.

\newpage

\noindent{\bf{VIII. Matter Distribution}}

\medskip

Because standpoint trajectories control Hubble-scale flow of matter, it is
meaningful to speak of a ``distribution of trajectories.''
Lorentz invariance of Hubble-scale distribution in infinite-age spacetime
constitutes model definition of ``homogeneous universe''.
{}From a selected standpoint to which zero initial rapidity is assigned,
Lorentz
invariance means {\it other } standpoints of initial rapidity $\vec{\Delta}$
are
isotropically distributed and, in {\it{ magnitude}} of initial rapidity, have a
distribution proportional to
$$
\sinh^2\Delta d\Delta = {z^2(1+{z\over 2})^2\over  (1+z)^3} dz ,\eqno(VIII.1)
$$
once again in agreement with standard model (Kolb, Turner 1990)
for zero deceleration.
Normalized to (IV.8) at $\Delta =0$, interpretation may be made of (VIII.1) as
Hubble-scale ``matter distribution''.
Close $(z \approx \Delta \ll 1)$ to the selected standpoint where
$\vec{r}_{\bf{R}}\approx \tau_{\bf{R}}\vec{\Delta}$, such a distribution is
uniform in the
usual sense of density independent of location, but as
$\Delta\to\infty$ the density implied by (VIII.1) increases without limit.
The distribution is nonintegrable, corresponding to an ``infinite classical
universe'' as in Milne's kinematic cosmology.

Notice on the other hand that according to our luminosity distance (VII.11),
sources with age-independent spectrum and brightness proportional to mass would
mean an average luminosity of the sky distributed in $\Delta$ (or $z$)
according
to
$$
e^{-2\Delta} d\Delta = {dz\over (1+z)^3}.\eqno(VIII.2)
$$
Most observed light thus would originate at $z\ltap 1$.
(The simple form (VIII.2), by virtue of ignoring variation of average intrinsic
source brightness with age of source, is not to be regarded as a falsifiable
model prediction).

The {\it {quantum}} lower limit on classical age, $\tau_{min} \sim 10^{-9}
sec,$
 in principle keeps finite a standpoint's universe.
On the standpoint backward light cone the age of matter is
$e^{-\Delta}\tau_{\bf{R}}$,
so the lower age limit places a corresponding upper limit on $\Delta $ (or
$z$):
$\Delta_{max} \sim \ln \tau_{\bf{R}}/\tau_{min} (z_{max} \sim
\tau_{\bf{R}}/\tau_{min})$.
In practice a far smaller bound to the visible universe is erected
by observational impediments.
A maximum observable redshift from our present standpoint is  $z_{dec} \sim
1400$,
corresponding to the ``decoupling'' temperature (see Peebles 1993 and Section
X below) above which photon mean free path becomes small on standpoint scale.
Nevertheless, over
the ``observable'' interval $z < z_{dec}$ the standard model
with deceleration-parameter $q_0 \sim 1/2$ predicts matter distribution
in {\it redshift} increasing far  less rapidly with $z$ than that given by
(VIII.1).
Matter distribution provides potentially unambiguous model discrimination
(Peebles 1993).
\newpage
\noindent{\bf IX. Poincar\'e Symmetry in Standpoint Neighborhood.}

This brief section makes explicit the sense in which standpoint cosmology is
compatible with ``usual'' classical physics inside any homogeneous-universe
neighborhood that is small on Hubble scale.
Consider two standpoint coordinate systems labeled by
$$
(R, \vec{\beta}) \ \hbox{and}\ (R', \vec{\beta}') \ \hbox{with}\ \abs{R-R'}
\ll \half (R +R')\ \hbox{and}\ \abs{\vec{\beta}}\ll 1, \abs{\vec{\beta}'}\ll 1.
$$
``Neighborhoods of standpoints'', defined loosely by
$$
\eqalignno{
\abs{ t-R} \ll R, \ &\abs{\vec{r}} \ll R,\cr
\abs{ t' -R'} \ll R', &\abs{\vec{r}\; '} \ll R',&(IX.1)\cr}
$$
then map onto each other even though the {\it{full}} spacetimes do {\it {not}}
map. (For example, because the maximum possible invariant distances from big
bang within these spacetimes  are $4R$ and $4R'$, respectively, if $R' > R$ the
portion of ${\bf{R}}'$ spacetime near abyss where $4R < s < 4R'$ does not map
onto ${\bf{R}}$ spacetime.) Employing the strategy (III.3) we may ask for the
relation between corresponding points $(t, \vec{r})$ and $(t', \vec{r}\; ')$
within
the neighborhood (IX.1).
One finds
$$
\eqalignno{
(t-R) &- (t'-R') \approx \tau'-\tau,\cr
\vec{r}& - \vec{r}\; ' \approx (\vec{\beta}'-\vec{\beta})
{\tau + \tau '\over 2},&(IX.2)\cr}
$$
up to corrections of order $(R+R')^{-1}$.
Change of standpoint is thus equivalent to a familiar Poincar\'e displacement.
Adding the consideration that, to the foregoing order, metric is Minkowskian
within the neighborhood (IX.1) for both coordinate systems, one recognizes
usual
Poincar\'e covariance of a unique noncompact spacetime.
For physics within this neighborhood any Poincar\'e transformation
may be invoked such that errors due to finiteness of universe are tolerable.

\newpage

\noindent{\bf{X. Thermodynamic Approximation?}}

\medskip
Because energy density varies inversely with square of age,
near sufficiently young standpoints one expects particle mean free path (in
time
as well as in space) to become small on the scale of $R$, allowing a
thermodynamic
approximation to develop meaning.
Isotropy of universe as viewed from a standpoint (using standpoint coordinates)
makes natural an association with young standpoints of temperature and
pressure,
as well as energy density; expectation is that such quantities will be found
in homogeneous-universe approximation to
be dependent only on standpoint age.
Accompanying energy density $\sim {1\over GR^2}$, a temperature
monotonically decreasing with standpoint age is anticipated.
Not yet under control, however, is the thermodynamic role of gravity. Model
geodesics imply unambiguous gravity and we have seen how attractive
gravitational forces provide universe ``confinement''- -defining a
spatially-spherical spacetime ``box'' of radius $R/2$.
But details of this ``box'' are
unorthodox to a degree that momentarily is frustrating effort to formulate a
consistent thermodynamic approximation.

Assuming thermodynamic equilibrium for sufficiently-small
standpoint ages, with
radiation decoupling as temperature at a certain age falls below atomic
ionization energies, a thermal photon (``black-body'') spectrum would survive
to
later ages with ``photon temperature'' decreasing inversely with age.
The energies of all decoupled photons diminish by a common factor as age
advances.

Nucleosynthesis must occur near standpoints whose temperature allows nuclear
reactions but, in absence of thermodynamic gravity understanding, calculations
have not yet been attempted.
It is momentarily unknown what standpoint cosmology predicts for light-element
abundances generated by primordial nucleosynthesis.
Making a preliminary crude guess that, during thermodynamic equilibrium,
energy density varies as $T^4$, the ratio $\sim 10^{22}$ is expected  between
age of photon
decoupling and the minimum classically-meaningful age, $\tau_{min}$.
The latter  accompanies a
maximum classically-meaningful temperature near  TeV scale.
The MeV-scale temperatures needed for nucleosynthesis would occur
 near an age $\sim 10^{10} \tau_{min} \sim 10 \ sec$.
\newpage

\noindent{\bf{XI. Concluding Remarks}}

\medskip

Not described here but presented in a separate paper (Chew 1995) is
standpoint-cosmology prescription for ``weak'' gravity - - small departures
from
Minkowskian metric at standpoint, departures generated by
matter-distribution {\it inhomogeneities} much less potent than black holes.
For ``weak'' inhomogeneities characterized by length scales well below Hubble
scale, the standpoint prescription concurs with Einstein theory.
Only for inhomogeneity scale approaching Hubble might there be significant
difference.
Almost all previous work on gravity-induced fluctuations in matter density is
sustained (see Peebles 1993).

For {\it any} matter distribution generating {\it large} metric deviation from
Minkowskian form, the new model's non-Riemannian structure will generate
unorthodox predictions.
Up to present, however, exploration of these predictions has been confined to
``homogeneous-universe'' calculations of Hubble-scale metric curvature - -
calculations reported in the present paper.
Investigation of small-scale ``strong'' inhomogeneities (``black holes'')
remains for the future.

Prime candidate for early falsifier of standpoint cosmology is the predicted
redshift dependence of luminosity distance (VII.1), but determination of matter
distribution up to redshifts
$\gtap .5$ could quickly eliminate the new model.
Although ability of the new model to explain light-element abundances is not
yet
established, cosmic background radiation presents no qualitative challenge.

Motivation behind classical standpoint cosmology has been, not addition
of curvature
to Milne's 1935 kinematic cosmology, but rather representation of the
symmetry of an underlying quantum model.
That symmetry implies for each standpoint a quartically-metricized {\it
compact} spacetime.
The  compactness in turn requires classical curvature:
Out of quantum {\it symmetry} has flowed classical
{\it dynamics}.

In homogeneous-universe approximation the quartic metric has yielded
the geodesics described in the present paper, which for {\it infinite-age}
standpoints reduce to those of Milne - - a limit where
all standpoint spacetimes become isomorphic to each other, noncompact and
Minkowskian.
Although {\it physical} spacetime is curved, belonging to a {\it finite}-age
standpoint,
the following  striking set of redshift-related
phenomenological features from Milne's model have survived in standpoint
cosmology:

(A) ``Age of universe'' equals a Hubble time defined by redshift.

(B) Luminosity distance and angular-diameter distance depend
 on redshift in the manner characterized standardly as
``zero deceleration'' (despite nonvanishing curvature of standpoint spacetime).

(C) The entire universe is in principle observable from any standpoint, with a
nonintegrable distribution in redshift that is uniquely determined by Lorentz
symmetry.

(D) Redshift factor equals ratio of observer age to source age (although total
redshift combines Doppler and gravitational shifts).

Even if redshift-expressible predictions of standpoint cosmology all turn out
indistinguishable from those of Milne's kinematic
model, geometrical features differ.
For example, at given age, the radius of a standpoint universe is larger
than that of Milne by a factor $\half + {1\over \sqrt{2}}$,
and the ratio of distance to Hubble-flow velocity is larger by a factor
$1+{1\over
\sqrt{2}}$.
Despite observational impracticability of investigating the foregoing subtle
differences,
experimental determination of mean energy density is widely regarded
feasible, and here Milne's model (unacceptably) seems to imply
$\Omega_0=0$, while standpoint cosmology predicts
$\Omega_0 = 2-\sqrt{2}$.

The current competition, of course, is not with kinematic cosmology but with a
``standard'' cosmology based on Einstein's theory of
gravitation.
Because the latter was originally formulated without regard for quantum
principles and without
regard for meaninglessness of time ``before big bang'', its reliability
at Hubble scale  may be
questioned.
Phenomenologically-viable alternatives should not be ignored, especially if
they entail fewer
arbitrary parameters.
A useful although not understood mnemonic is that, apart
from energy density, all predictions listed here coincide with
zero-deceleration standard-model predictions.

\medskip

\noindent{\bf{Acknowledgment}}

Suggestions from P. Eberhard, J. Finkelstein, S. Perlmutter, and H.P. Stapp are
gratefully acknowledged.
These colleagues have steered the author away from a variety of model
misinterpretations.
This work was supported by the Director, Office of Energy
Research, Office of High Energy and Nuclear Physics, Division of High
Energy Physics of the U.S. Department of Energy under Contract
DE-AC03-76SF00098.

\newpage

\noindent{\bf{Appendix. Gravitational Redshift Along Standpoint Light Cone.}}

\medskip

Differentiating Formulas (IV.11a) and (IV.11b) and taking the quotient leads to
the following expression for particle velocity as it varies along a
{\it{radial}} standpoint-spacetime geodesic:
$$
v_{\bf{R}} \equiv {dr_{\bf{R}}\over dt_{\bf{R}}}= { b(1-{t_\infty\over 4R}) -
{r_\infty\over 4R}\over 1 - {t_\infty\over 4R} - b{r_\infty\over 4R}
}.\eqno(A.1)
$$
The constant $b$, limited to the interval $-1\leq b\leq +1$, is the
radial-motion special case of the 3-vector $\vec{b}$ appearing in the general
geodesic (IV.12).
Notice from (A.1) that in each of the two limits, $b\to \pm 1, \ v_{\bf{R}}$
approaches the same limit as $b$ and that, for any allowed $b$,
$\abs{v_{\bf{R}}} \leq 1$.
(It may also be verified that if $\abs{b} <1$ then, along the abyss boundary,
where $t_\infty + r_\infty =4R$, one {\it always} finds $v_R = -1$ - - i.e.,
{\it{inward}} matter motion {\it{parallel}} to boundary as required by
confinement to compact standpoint spacetime.)

Light arriving at standpoint from a source located on standpoint backward light
cone corresponds to the limiting case $v_{\bf{R}} =b=-1$.
Our deduction of gravitational redshift will invoke the relation
$$
v_{\bf{R}} = {e^{\gamma_{\bf{R}}} - e^{-\gamma_{\bf{R}}}\over
e^{\gamma_{\bf{R}}}+ e^{-\gamma_{\bf{R}}}}\eqno(A.2)
$$
between particle velocity $v_{\bf{R}}$ and particle {\it rapidity}
$\gamma_{\bf{R}}$.
Even in the limits $v_{\bf{R}}\to \pm 1$, where $\gamma_{\bf{R}}\to\pm\infty$,
there is (finite) {\it rapidity variation} along the trajectory- -
corresponding
to energy shift.
For zero-mass particles, energy varies in proportion to
$e^{\delta\abs{\gamma_{\bf{R}}}}$
where $\delta\abs{\gamma_{\bf{R}}}$ means {\it change} in the
absolute value of $\gamma_{\bf{R}}$.
We may alter (A.1) to a rapidity-variation relation, applicable to incoming
photons, by asymptotically expanding (A.2) for large negative rapidity,
$$
\eqalignno{
 v_{\bf{R}} &= - 1 + 2 e^{2\gamma_{\bf{R}}} +\ {\hbox{terms of order}} \
e^{+4\gamma_{\bf{R}}}\cr
&- \gamma_{\bf{R}} \gg 1&(A.3)\cr}
$$
and making a corresponding expansion of (A.1) around $b=-1$.
Writing $b=- 1+\epsilon, \ \epsilon > 0$, one finds
$$
v_{\bf{R}} =   { -1+\epsilon\ { 4R-t_{\infty} \over 4R -(t_\infty - r_\infty)}
\over 1-\epsilon\ { r_\infty\over 4R -(t_\infty - r_\infty)}}
 {\mathop{=}_{\epsilon \ll 1} }
-1+\epsilon { 4R - (t_\infty + r_\infty)\over
4R -(t_\infty - t_\infty)} + \hbox{terms of order} \ \epsilon^2.\eqno(A.4)
$$
By comparing (A.3) with (A.4), it may be inferred that
$$
e^{2\gamma_{\bf{R}} }\approx {\epsilon\over 2}{4R -(t_\infty + r_\infty) \over
4R - (t_\infty - r_\infty) },\eqno(A.5)
$$
for $\epsilon \ll 1, -\gamma_{\bf{R}} \gg 1.$
Because $\gamma_{\bf{R}}$ is negative, photon energy
is thus proportional to
$$
\left[ {4R-(t_\infty - r_\infty)\over 4R - (t_{{\infty}}+
       r_\infty)}
\right]^{1/2} =
\left[ {2R-(t_{\bf{R}}-r_{\bf{R}})\over 2R-(t_{\bf{R}}+r_{\bf{R}})}\right]
^{1/4},\eqno(A.6)
$$
the right-hand form following if we remember from (IV.10) that
$$
1-{t_\infty \pm r_{\infty}\over 4R} = \left[ 1-{t_{\bf{R}}\pm r_{\bf{R}}\over
2R} \right]^{1/2}.\eqno(A.7)
$$
Along the standpoint backward light cone, $t_{\bf{R}} + r_{\bf{R}} = R$, while
at standpoint $r_{\bf{R}}=0$.
It follows from (A.6) that the gravitational redshift factor for
propagation between $r_{\bf{R}} =r_{s}$ and standpoint is
$$
\left( 1+ {2r_s\over R}\right)^{1/4}.\eqno(A.8)
$$

Of interest in principle (although not in practice) is gravitational redshift
of
light {\it{ emitted}} from standpoint and proceeding along the standpoint's
{\it{forward}}  light cone where $t_{\bf{R}} - r_{\bf{R}} =R$. Repeating the
foregoing calculation for the limit $v_{\bf{R}}\to + 1, \gamma_{\bf{R}}\to +
\infty$, one finds gravitational energy-reduction during propagation in
${\bf{R}}$ standpoint system by a factor
$$
\left( 1 - {2r_s\over R}\right)^{1/4}\eqno(A.9)
$$
for light reaching a distance $r_s$ from standpoint.
At abyss, where $r_s = R/2$, all ${\bf{R}}$-system photon energies are thus
reduced to zero.
On the other hand, if one thinks of light {\it absorption} by matter at and
moving with some {\it{other}} standpoint located on forward cone, the motion of
that standpoint produces a Doppler shift so that the {\it net} redshift in the
usual sense continues to be given by the ratio of standpoint ages.
\newpage

\noindent{\bf{References}}

\noindent Chew, G.F. 1994, {\it Lawrence Berkeley Laboratory preprint
LBL-36314},
    to be published in  Foundations of Physics.

\noindent Chew, G.F. 1995, {\it Lawrence Berkeley Laboratory preprint
LBL-36809},
    submitted to the  Astrophysical Journal.

\noindent Kolb, E.W. and Turner, M.S. 1990, {\it The Early Universe},
Addison-Wesley.

\noindent Milne, E.A. 1935, {\it Relativity, Gravitation and World Structure},
    Oxford: Clarendon Press.

\noindent Peebles, P.J.E. 1993, {\it Principles of Physical Cosmology},
Princeton University Press.

\noindent Weinberg, S. 1972, {\it Gravitation and Cosmology}, Wiley, N.Y.

\end{document}